\documentclass[letterpaper]{article} 
\usepackage{aaai23}  
\usepackage{times}  
\usepackage{helvet}  
\usepackage{courier}  
\usepackage[hyphens]{url}  
\usepackage{graphicx} 
\usepackage{natbib}  
\usepackage{caption} 
\usepackage{booktabs}
\usepackage{caption}
\usepackage{subcaption}
\usepackage{algorithm}
\usepackage{algorithmic}

\usepackage{newfloat}
\usepackage{listings}
\DeclareCaptionStyle{ruled}{labelfont=normalfont,labelsep=colon,strut=off} 
\floatstyle{ruled}
\newfloat{listing}{tb}{lst}{}
\floatname{listing}{Listing}
\title{Comfort Foods and Community Connectedness: \\Investigating Diet Change during COVID-19 Using YouTube Videos on Twitter} 
\author{Yelena Mejova\thanks{Authors contributed equally to this work}$^1$, Lydia Manikonda\footnotemark[1]$^2$ }
\affiliations{
$^1$ ISI Foundation\\
$^2$ Rensselaer Polytechnic Institute\\
yelenamejova@acm.org, manikl@rpi.edu }

\begin{document}

\maketitle

\begin{abstract}

Unprecedented lockdowns at the start of the COVID-19 pandemic have drastically changed the routines of millions of people, potentially impacting important health-related behaviors.
In this study, we use YouTube videos embedded in tweets about diet, exercise and fitness posted before and during COVID-19 to investigate the influence of the pandemic lockdowns on diet and nutrition. In particular, we examine the nutritional profile of the foods mentioned in the transcript, description and title of each video in terms of six macronutrients (protein, energy, fat, sodium, sugar, and saturated fat). These macronutrient values were further linked to demographics to assess if there are specific effects on those potentially having insufficient access to healthy sources of food. 
Interrupted time series analysis revealed a considerable shift in the aggregated macronutrient scores before and during COVID-19. In particular, whereas areas with lower incomes showed decrease in energy, fat, and saturated fat, those with higher percentage of African Americans showed an elevation in sodium. Word2Vec word similarities and odds ratio analysis suggested a shift from popular diets and lifestyle bloggers before the lockdowns to the interest in a variety of healthy foods, communal sharing of quick and easy recipes, as well as a new emphasis on comfort foods. 
To the best of our knowledge, this work is novel in terms of linking attention signals in tweets, content of videos, their nutrients profile, and aggregate demographics of the users. 
The insights made possible by this combination of resources are important for monitoring the secondary health effects of social distancing, and informing social programs designed to alleviate these effects. 

\end{abstract}

\section{Introduction}

During the outbreak of COVID-19 pandemic, global social distancing restrictions forced millions of people to stay at home. 
As daily activities and dietary habits changed, surveys show that cooking at home, and physical inactivity alongside the imposed seclusion has increased \cite{bennett2021impact}.
However, surveys disagree whether the consumption of snack foods and alcohol has gone up or down, and it is possible that different subpopulations may be responsible for either shift.
In this context, social media played a key role in connecting with others online, sharing information, and discussing experiences. 
One of the most common ways of making such posts in this context is to include URLs from other platforms, such as YouTube, which hosts videos about recipes, exercise routines and videos around diet and health. 
We ask, then, \emph{what shifts in dietary preferences around the COVID-19 lockdowns are evident in the online content sharing?}
Existing studies~\cite{abbar2015you,fried2014analyzing,mejova2015foodporn} have considered different features of language posted online when sharing about foods to characterize dietary choices and nutrition. 
We extend this literature by considering fine-grained nutrition information at a macronutrient level to track the change in diet behaviors of various subpopulations during the onset of COVID-19 restrictive measures. 

We contribute a dataset of tweets about diet, nutrition and fitness shared publicly \emph{before} (between June 19, 2019 to March 15, 2020) and \emph{during} (March 16, 2020 to June 24, 2020) COVID-19 pandemic that contain YouTube URLs. 
Using the content of the videos and the context of the tweets in which they were shared, we investigate the dietary preference shifts in this specific time period. 
Further, we focus on populations likely to have worse access to healthy and affordable food, including those living in lower income locations or being a vulnerable minority. 
In particular, we consider the following macronutrients: \emph{protein}, \emph{fat}, \emph{sugar}, \emph{sodium}, \emph{energy}, and \emph{saturated fat} from the metadata of the videos whose URLs were embedded in the tweets. 
We then explore the shifts in the context around food-related content in order to reveal a shift in focus, especially around (1) dieting and (2) comfort foods.


Specifically, we provide insight about the following research questions:
\begin{itemize}
    \item \emph{Has the macronutrient quality of mentioned food improve overall, and especially for disadvantaged demographic groups?} Our investigation revealed that there is a small but significant shift in energy, fat, saturated fat, and sugar, all in the negative direction, suggesting an overall shift in focus to healthier foods. On the other hand, we found an increase in sodium-laden foods in counties with higher African American population, but a decrease in energy, fat, and saturated fat in counties with lower household income.
    \item \emph{Has the interest in dieting increased during the pandemic?} Using shifts in embeddings of food-related words in videos we found ``healthy life'' to be increasingly associated with ``veganfood'' and ``fullyraw'', shifting away from celebrity-driven wellness content. However mentions of unhealthy food restriction became more prominent with ``emotional'' and ``binge'' association with ``mindful'' keyword. 
    \item \emph{Has the mentioning of comfort foods increased during the pandemic?} Unlike in videos, the context provided in the tweets includes both mentions of diet restrictions like ``glutenfree'', but also includes comfort food such as ``waffles'', ``treat'', and ``dough'', and especially towards breakfast foods in the lower-income areas. Thus the salient foods emphasized in the tweets include both shifts toward dieting and comfort foods, albeit in different populations. Finally, the expected shifts toward homecooking were also observed.
\end{itemize}

In this work, we illustrate the richness of content the multimedia links shared on Twitter may provide in order to both quantitatively and qualitatively examine population health-related behaviors around the unprecedented events of the first COVID-19 lockdowns. 

\section{Background and Related Work}


\subsubsection{Food Consumption during Disasters}
There are existing studies on how stress can affect the brain's response to high calories food cues and predispose to obesogenic eating habits~\cite{fowles2012stress,michels2012stress,tryon2013chronic}. Given the focus of this research on the influence of a pandemic, we aim to consider literature surrounding food consumption during disasters. It has been shown that eating behaviors change under acute life stressors, such as war and natural disasters~\cite{ho2002lifestyle,mcintosh1980food}. 
Research shows that people change their eating habits by either reducing the number of meals eaten or go whole days without eating; or rely on less preferred and less expensive food~\cite{israel2012impacts,mckenzie2005buying}. Especially during the strict lockdowns and social distancing guidelines, existing literature states that there is a slight increased physical activity with a higher adherence to healthy diets~\cite{christofaro2021physical,di2020eating,teixeira2021eating}. A study conducted during the initial stage of COVID-19 lockdown in China showed that staying at home or working from home was associated with an increase in animal products, vegetable, fruit, mushroom, nut, water and snack intake~\cite{yang2021eating}. Contradicting these studies, a survey conducted by \cite{esobi2020impact} showed that limited access to food due to restricted store hours and inventory shortages led to an increased intake of fast foods and packaged foods. Also, the same study highlights an increased consumption of fruits and vegetables that has been attributed to an increase in antioxidant status resulting in weight loss and increased resistance to the disease~\cite{mattioli2020covid}. Relating to this existing literature, our proposed work not just focuses on the trends of food types or categories during COVID-19 but also on potentially disadvantaged demographics by using a large-scale dataset. 
This is especially urgent, since recent study in Columbus, Ohio~\cite{kar2021covid} found that a higher percentage of low-income customers was associated with lower store visits during the lockdown period, and that stores with a higher percentage of white customers declined the least and recovered faster during the reopening phase.
Here, we focus both on income-based and racial areas when examining changes in food content sharing.

\subsubsection{Studying Food Consumption on Internet}
Web and social media platforms have been actively used to study health-related behaviors around nutrition and other food-related patterns~\cite{abbar2015you,fried2014analyzing,mejova2015foodporn,vydiswaran2020uncovering}. Twitter has been used to study diverse questions including: 1) neighborhood food environment~\cite{nguyen2017social}, 2) relationship between obesity rates and geo-coded Twitter posts related to food intake~\cite{gore2015you}, 3) contexts of food choices~\cite{vidal2015using}, 4) topics posted and discussed related to diet, diabetes, exercise and obesity~\cite{karami2018characterizing}, etc. 
Despite the many biases of self-expression on social media, foods mentioned online have been shown to correspond to offline health statistics~\cite{min2019survey}, pointing to its usefulness in now-casting health-related behaviors.
Twitter has also been used as surveillance for individuals who track their food intake to examine the relationship between dietary and behavioral factors~\cite{hingle2013collection}. 
Researchers also looked at other platforms such as Instagram to characterize poor access to healthy and affordable food as well as perceptions and tracking about the characteristics of food~\cite{chung2017personal,ofli2017saki}. 
In particular, those living in ``food deserts'' post foods higher in fat, sugar and cholesterol by 5-17\% over those in similar, but non-``desert'' locales~\cite{de2016characterizing}.
Furthermore, there is a small set of literature focusing on algorithmic nutritional estimation to judge whether a recipe is healthy using various cues associated with the data~\cite{lee2019estimating,rokicki2018impact}.
For example, \citet{trattner2017investigating} use 7 macronutrients (6 we also consider, plus fiber) to estimate healthiness of recipes from Allrecipes.com.
However, much of this existing literature focuses on using the post-related content such as text or images to evaluate healthiness of foods or their authors. 
Our proposed work instead utilizes rich metadata around the media cited in the tweets including transcript, description, title, etc., to compute macronutrient estimation. Moreover, our proposed work evaluates the semantic and syntactic changes of word associations by representing both the video content as well as tweet content separately in an embedding space. 

\subsubsection{Sharing YouTube URLs on Twitter}
Twitter is particularly known for sharing and consuming news~\cite{kwak2010twitter,morgan2013news} and topics around politics, product reviews, food reviews, customer experiences, etc~\cite{manikonda2016tweeting}. 
Studies~\cite{abisheva2014watches,christodoulou2012youtubediffusiontwitter} have shown that Twitter helps diffuse YouTube content via embedded links help them gain viewership. 
Recently, a study of YouTube links shared on Twitter has shown that such cross-platform posting was affected by the lockdown measures in countries around the world, and that it negatively correlates with mobility \cite{mejova2021youtubing}. Complementing the existing work, our proposed approach evaluates the influence of COVID-19 on shifts in macronutrient values and food-related vocabulary usage using not just tweet content but also video content.

\section{Data Collection and Preprocessing}

We begin by querying the Twitter Streaming API for ``youtube'' and ``youtu.be'' to collect tweets sharing YouTube videos. We then compile a set of keywords related to diet and nutrition and use them to filter the captured tweets\footnote{for a full list see https://tinyurl.com/dietkeys}. We have collected 249,051 such tweets from June $19^{th}$ 2019 to June $24^{th}$ 2020. The main reason to consider this time period is to take into consideration the period before COVID-19 was officially declared as a pandemic and during the first lockdowns of the pandemic. There is a total of 112,936 unique urls extracted from these tweets. Using a python library \emph{pytube}, we extracted the metadata elements of each YouTube URL including \emph{title}, \emph{description}, \emph{author}, \emph{length} and \emph{rating} (computed using the ratio of likes and dislikes and scaling that value to lie between 0 and 5, with 0 being low and 5 being high). Additionally, for each YouTube URL we extracted the available \emph{transcript} in the ``English" language. After extracting the metadata of each YouTube URL, we have a total of 40,098 URLs that have all their titles, descriptions and transcripts downloaded. 

\subsubsection{Demographics Data}
We use the county-level census data that was made available publicly by the County Health Rankings \& Roadmaps (\url{https://www.countyhealthrankings.org/}), from the University of Wisconsin Population Health Institute. Since the Twitter data we utilize in this paper is from 2019 to 2020, we use the county health ranking data from 2019. We use the FIPS (Federal Information Processing Standards) code of the counties to map the demographics to the tweet location. 

\section{Methods}

\begin{figure}[t]
\includegraphics[width=\columnwidth]{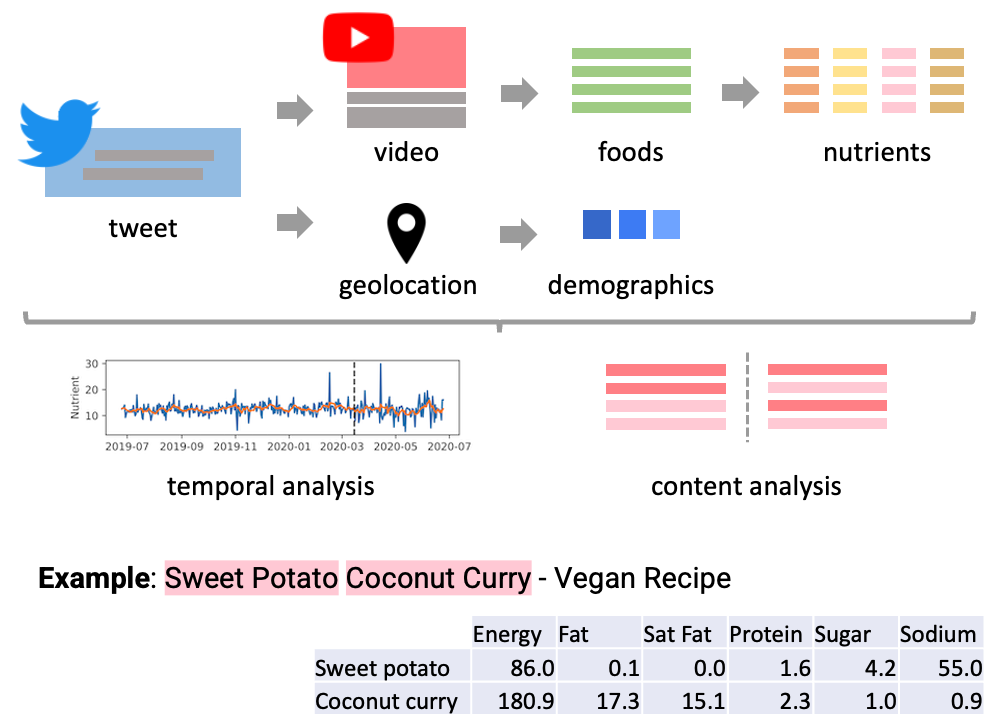}
\centering
\caption{Diagram representing the data processing pipeline. An example title of a video is shown, with detected foods highlighted and nutrients listed.}
\label{fig:datapipeline}
\end{figure}

We begin by processing the URLs shared in the tweets to identify whether they are related to nutrition or dietetics (see Figure~\ref{fig:datapipeline}). Metadata for these relevant URLs are extracted and used to infer the values of macronutrients. 
Specifically, we consider \emph{Protein}, \emph{Fat}, \emph{Sugar}, \emph{Sodium}, \emph{Energy}, and \emph{Saturated Fat} as the main set of nutrients for this research (all measured in grams, except sodium, which is measured in milligrams). 
For each URL of the YouTube video, we have the estimated nutrients as well as the related metadata which includes \emph{title of the video}, \emph{views}, \emph{rating}, \emph{description of the video}, and \emph{video transcript}. We then consider the Twitter profiles of users who shared the video URLs in our database and link them with the demographics data.

\subsection{Processing Diet-related URLs and Extracting Video Metadata}
We extract the video metadata using \emph{pytube} \url{(https://pypi.org/project/pytube/)} library. For each URL, we download their corresponding \emph{title}, \emph{description}, \emph{author}, \emph{length} and \emph{rating}, and \emph{video transcript} (only in english language, due to library limitation). After this step, we have a total of 97,314 URLs with a downloaded title, 94,037 URLs with a downloaded description, and 54,631 URLs with a downloaded transcript. 
All these three metadata elements are used in estimating nutrients for further analysis described in this paper.

\subsection{Identifying Food-related Videos through Classification}
Upon manual examination, we find that not all videos are about food, but some general wellness and completely unrelated videos passed the filter.
We use a two-step process to refine the relevance of the videos. 
First we use a lexicon dictionary \cite{salvador2017learning} to find if a given video has at least one food ingredient present in any of their metadata fields, resulting in 40,098 URLs. 
Second, we perform classification using the trained model on manually labelled videos. 
We then randomly select 300 videos 
and manually label them as a \emph{food} or \emph{non-food} category.
The annotation was performed mostly by one author with discussion over unclear cases amongst all authors. An annotator agreement exercise over 30 videos resulted in 2 disagreements, or a Cohen’s Kappa of 0.88.
The guidelines outlined the definition of video being ``food-related'' when the main topic of the video is a recipe or a particular ingredient. For example, a video on the physical mechanics of ketosis during keto diet is labeled as not food-related, even if it mentions food, whereas a keto-friendly recipe is labeled as food-related.
We use this as training data by first extracting the BERT sentence embeddings for each video using their metadata (combining title, description and transcript). 
We train three different classifiers -- 1) Random Forests, 2) Gradient Boosting, and 3) Multi-layer Perceptron (all three using the sklearn library of Python).
Evaluating the models using ROC AUC on 5-fold cross validation, we find that Random Forests classifier generates most accurate results with AUC = 0.97 (average F1-score = 0.81, precision = 0.75 and recall = 0.88). We then use the trained classifier to label the remaining URLs to obtain the final class labels, resulting in 20,455 videos (51\% of filtered set) are about food. 


\subsection{Estimating Nutrients from the Video Metadata}

We leverage a lexicon that combines basic ingredients and dishes that contains 21,162 entries alongside their nutritional value to annotate the retrieved videos compiled from previous public health nutrition research \cite{salvador2017learning,alajajian2017lexicocalorimeter}
(using title, description and transcript text). 
We extract the unique set of keywords as the corresponding vocabulary from the metadata elements while preserving the order in which these keywords are originally mentioned. 
We then manually remove ingredients which are likely to be mentioned not as nutrients such as `water' (for example, ``body water''), `fat' (``body fat'', though we understand that there are recipes where it is suggested to use fat explicitly such as bacon fat), `essential oil', etc. 
Finally, we compute the average nutritional value over all detected foods in a video to summarize its nutritional profile. 
We manually verified this process by randomly sampling 30 different videos and extracting food-related vocabulary (ground truth). We compute the precision and recall values for each video as the percentage of keywords that were automatically extracted that matches with the manually extracted keywords and the percentage of keywords from manually extracted vocabulary matches with the automatically extracted keywords, respectively. This estimation provided an average scores of 0.93 and 0.94 for precision and recall values respectively, showing a reasonable performance. 


\subsection{Geolocation}
In order to associate the tweets with locations, we use the location field in the posting user's profile. Employing an approach from previous literature \cite{mejova2021youtubing}, we map the location fields to the Geonames geographical database\footnote{https://www.geonames.org/}, and select only those within the United States, which could be mapped to a county. To make sure most of the matches are correct, we manually examine the 1000 most frequently matched locations and remove spurious ones (for instance ``home'' or ``In Heaven''). Out of the tweets having macronutrient information, we are able to geolocate 10,781 tweets.



\subsection{Change in Nutritional Values}

To understand whether the nutritional content of the foods mentioned by the social media users changes after the onset of COVID-19, we employ Interrupted Time Series analysis (ITS)~\citep{bernal2017interrupted}, which aims to estimate the effect of an intervention which has a well-defined starting time.
We choose this model in order to focus on the change in the average each nutrition value found in the two time periods of the data.
Specifically, we examine the 7-day average of each of the six nutritional values detected above, in order to gauge the extent of the behavior change, and as the intervention we take March 15, 2020, as when state-wide lockdown measures started to take place\footnote{\url{https://www.pbs.org/newshour/politics/most-states-have-issued-stay-at-home-orders-but-enforcement-varies-widely}}.
We employ Ordinary Least Squares (OLS) regression to model the nutrient time series using two variables: $P$ signifying time passage in days and $X_t$, an indicator signifying the beginning of the intervention period (in our case, it is 1 after March 15, 2020).
For example, modeling change in the protein value of the posted foods after the onset of COVID-19, we use the following equation:
\[ y_t = \beta_0 + \beta_1 P + \beta_2 X_t + \beta_3 P X_t \]

\noindent where:
\begin{itemize}
    \item $y_t$ is the smoothed protein value at the time $t$
    \item $\beta_0$ is the baseline value at the beginning of the time series
    \item $\beta_1$ is the baseline change in value over time before the COVID-19 date
    \item $\beta_2$ is the change in value at the COVID-19 date
    \item $\beta_3$ is the trend (slope) change following the COVID-19 date
\end{itemize}

In this paper, we focus specifically on the $\beta_2$, the change in posting behavior around the beginning of the lockdowns. 
We also consider the p-value of the coefficient, considering $p<0.01$ as a statistically significant result.
As a baseline, we compare the p-values obtained in this excise to models run on a selection of 100 randomly chosen ``treatment'' dates in the period before the pandemic, while allowing for 90 days of data for before and after sub-periods. 
In particular, we compare how often changes of significance observed during the onset of COVID-19 happen before, intuitively providing a measure of how often significant changes in nutrition value happen in a ``normal'' setting. 
We consider the result to ``pass'' this baseline, if no shift was observed in these 100 trial with the same or greater significance.
We also check the data for seasonality by plotting the autocorrelation function, and for stationarity using the Dickey-Fuller test, finding the time series of all nutrients stationary at $p<0.05$.

\subsection{Change in Content}
Although the nutritional values of the content provides a valuable insight into the overall nutritional trends of the dataset, we continue the analysis by considering the textual content of the tweets and the videos they link to in order to understand the qualitative changes in people's diets and their self-expression around this topic.
We consider both the textual content around the video (title, description, and transcript), and the text of the tweet referring to the video, as representations of the content and context of the food sharing experience.
For both sources of text, we begin by tokenizing, lower-casing, and removing non-word characters, as well as English language stopwords.
We also remove mentions of other users, URLs, and the hashtag sign (but not the hashtag itself, as it often carries important information). 
Using this cleaned version of the text, we de-duplicate the data (effectively finding near-duplicates), thus preventing very popular content from taking over the frequency computations.
De-duplication was done on the ``cleaned'' text, thus making sure that small differences in formatting, addition of URLs, and stopwords did not constitute a new document.
Finally, we consider the tweets and videos shared before and during COVID-19 pandemic periods (March 15, 2020) as two distributions of words that can be compared.
For this comparison, we use two approaches. One is Odds Ratio that compares the relative frequencies of words, which is 1 if there is no difference in a word use before and during COVID-19, less than 1 if it is used less, and more than 1 if it used more.
Second is the cosine similarity of Word2Vec~\cite{mikolov2013distributed} embeddings of words related to `food', providing both the semantic and syntactic characteristics of words~\cite{devlin2018bert,vrehuuvrek2011gensim}. Please note that we are not looking for how a specific term or vocabulary evolved over time (and there is a plethora of work~\cite{kenter2015ad,liu2021statistically} which is supported by using statistical significance tests in terms of measuring whether the evolution of a word temporally is meaningful and significant). The goal is to mainly examine the significant differences between how words were associated with each other both semantically and syntactically in a qualitative manner.

\section{Results}

\subsection{Data Description}

As stated earlier, we first crawl Twitter to download appropriate tweets that have an embedded YouTube URL and are related to diet, nutrition and health. After cleaning the data over several steps of pre-processing, we finally obtain a total of 20,455 videos that are focused explicitly on food recipes. When we connect the URLs of these videos back to tweets, we found that on an average tweets (across the entire distribution, $min$=3004; $avg$=6823; $max$=9273) with food-related video urls were relatively similar in terms of volume before COVID-19 ($avg_{Before}$=6721.5) and during COVID-19 ($avg_{During}$=5371.0) with a $p$-value of 0.316 which means that even though means are slightly different, this difference is not statistically significant. However, the maximum number of tweets about food were posted in December 2019 before COVID-19 and in May 2020 during COVID-19. This may suggest that posting URLs embedded in tweets is a common practice and extending research to leverage such content will be valuable to investigate not only user behaviors but also their health and well-being.


We assess the suitability and reliability of the dataset to our further analyses by extracting top $n$-grams based on their frequency in the corpus. The top 10 \emph{uni}, \emph{bi}, and \emph{tri}-- grams (with stopwords removed) are shown in Table~\ref{tab:ngrams}. 
Indeed, the videos are focused on health and specifically diet. In a few instances, these videos are emphasizing on the type of diet such as ``keto'' or ``vegan'' as well as the health-related conditions such as ``type 2 diabetes'' or ``weight loss''. In other instances, phrases such as ``let know comments'', ``thank much watching'' primarily the speaker of the video engaging with their audience which is expected. 

\begin{table}[t]
\centering
\begin{tabular}{lll}
\toprule
\emph{uni}-grams & \emph{bi}-grams & \emph{tri}-grams \\
\hline
get & little bit & 1 2 cup\\
make & make sure & apple cider vinegar \\
right & weight loss & 1 4 cup\\
good & really good & plant based diet\\
little & plant based & want make sure\\
vegan & low carb & 1 2 tsp\\
people & keto diet & thank much watching\\
diet & olive oil & type 2 diabetes\\
food & lose weight & low carb diet\\
eat & let know & let know comments\\
keto & gluten free & low carb keto\\

\bottomrule
\end{tabular}
\caption{Top-10 $n$-grams where, $n \in $ \{1,2, 3\} extracted from metadata of videos }
\label{tab:ngrams}
\end{table}

Examining the most shared videos in either of the time periods, we find that these often come from online influencers and actors. 
In the period before the pandemic, the most shared video is actually not around health per se, but is a challenge on gaining weight by eating. 
It is followed by informational videos around veganism and the dairy industry. 
Only one out of the top 5 of pre-pandemic viral videos contains actual recipes, as a part of a video log of ``what I eat in a day''. 
The top videos during pandemic are instead more focused on recipes, topping with a ``fat bomb'' (keto-friendly), a beef bowl, and a nutritional supplement.  
These came both from professional chefs, and from internet personalities not showing any training in nutrition. 
Thus, in the most viral videos, we find a shift toward instructional content, and away from general and personal information.

\subsection{Macronutrients}
\begin{figure}[t]
\includegraphics[width=\columnwidth]{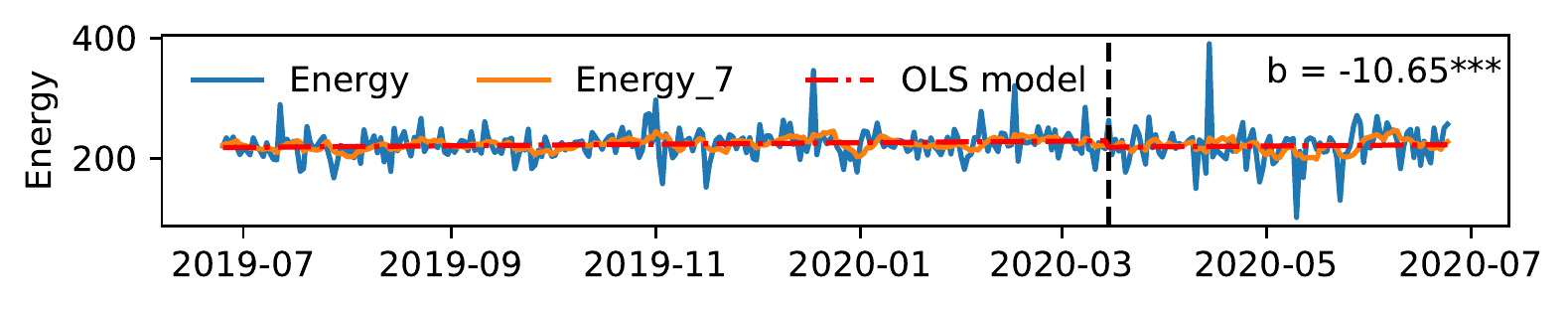}
\includegraphics[width=\columnwidth]{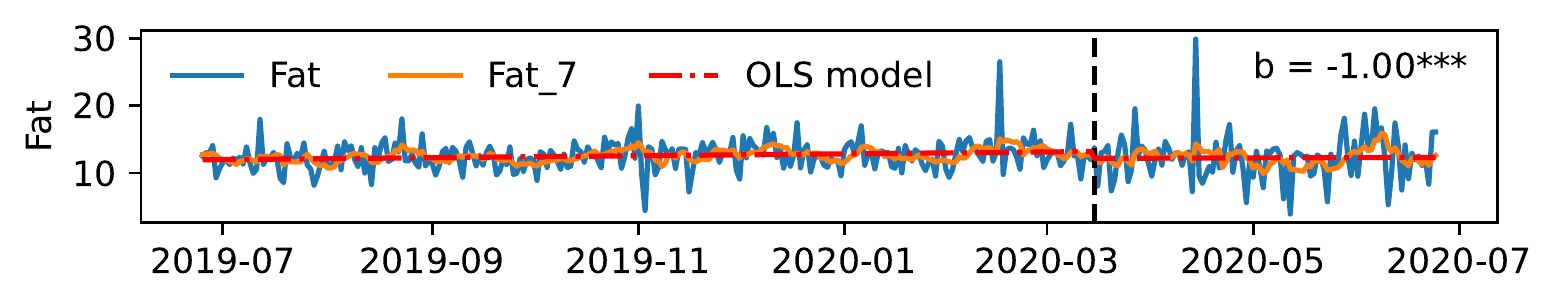}
\includegraphics[width=\columnwidth]{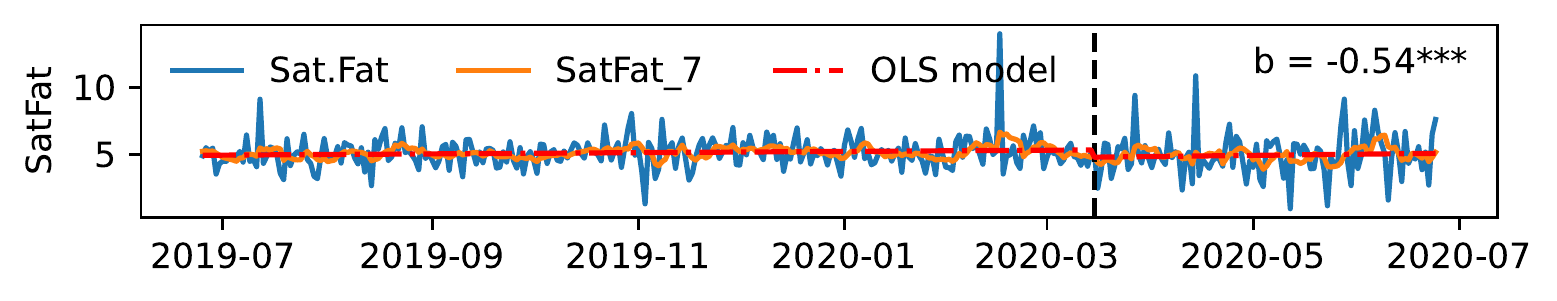}
\includegraphics[width=\columnwidth]{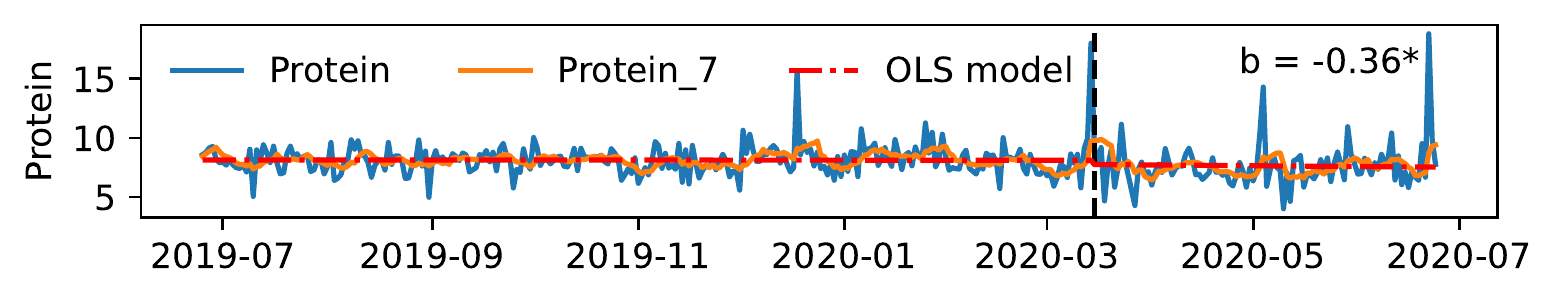}
\includegraphics[width=\columnwidth]{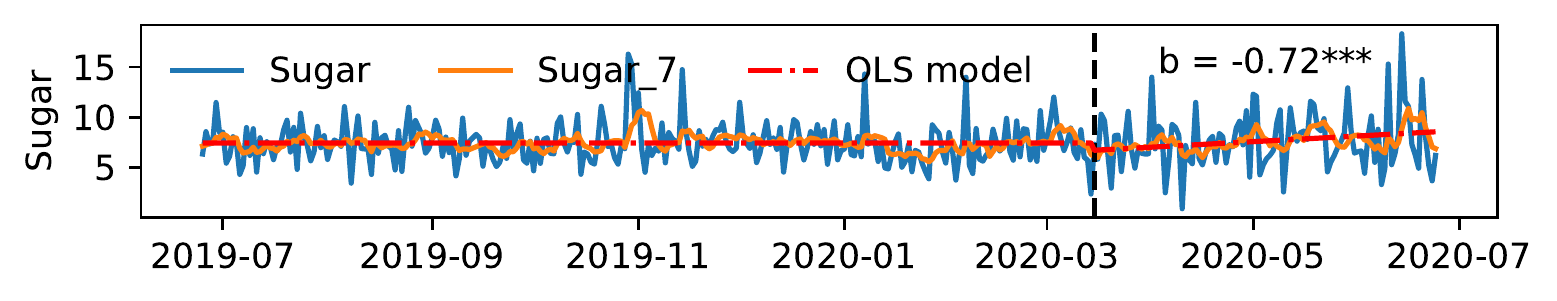}
\includegraphics[width=\columnwidth]{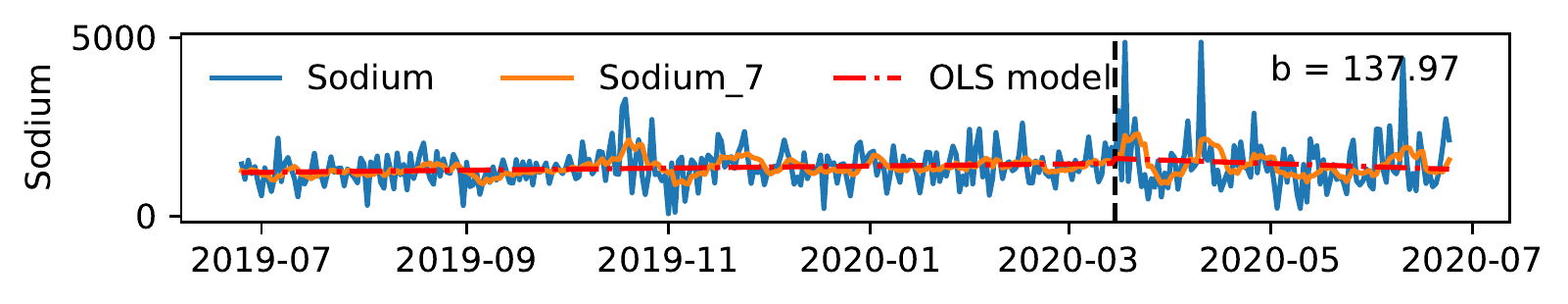}
\centering
\caption{Daily average nutrient value, 7 day average, and Interrupted Time Series model. In upper left, text shows the beta for lockdown variable (*** $p < 0.0001$, * $p < 0.01$).}
\label{fig:its_nutrients}
\end{figure}

In order to monitor the change in macronutrient content of the food mentioned in these videos, we create a time series for each and apply Interrupted Time Series (ITS) analysis. 
Figure \ref{fig:its_nutrients} shows the daily macronutrient values, the 7 day aggregates, and the ITS model. 
In the upper left, the text shows the $\beta$ coefficient associated with the change in the macronutrient value at the COVID-19 date, along with the associated $p$ value. 
We find that energy, fat, and saturated fat significantly decrease during the COVID-19 period, as well as sugar, which then increases over time (that is, $\beta_3$ is significantly positive at $p<0.001$). 
Instead, sodium shows an overall downward trend during the COVID-19 period (though note that this observation is sensitive to the duration of this period).
We find all results, except one for sugar, are uniquely significant when comparing to changes in nutrient scores during 100 randomly chosen breakpoints before COVID-19: that is, the p-values are smaller than for any change that occurred before COVID-19 period.
Note, however, that the change in sugar is still significant, but such significant changes were also likely to occur before COVID-19. 
Although significant, these changes are mostly small in aggregate, such as the change in energy of 10 calories, or 1 gram of fat.
We also note that the variability of the nutritional values increases during the lockdowns, compared to the previous period. 

\subsection{Macronutrients \& Demographics}


Analysis in Figure \ref{fig:its_nutrients} is computed over all users in our dataset, it is possible the effects may be different in subsets of the population.
Thus, we consider six demographic variables related to food access: income, education, percent African American, percent Hispanic residents, percent experiencing food insecurity, and percent having limited access to healthy food \cite{usda2020food}. 
In particular, we focus on the users that come from areas (identified at the level of a U.S. county) that are below or above the median (above, first two below and the last four above), for instance those coming from counties in with below median household income, or above median Hispanic residents.
We then perform the ITS analysis on the selected users and examine the changes in macronutrient values of the videos shared before and during the pandemic. 
We apply Bonferroni correction for multiple comparison to the $p$-values of the COVID-19 coefficient.
Additionally, we compare the resulting $p$-values to a random date baseline.
Table \ref{tab:nutrients_demogs} shows the significant changes in macronutrients for subsets of users ($p$-values passing random date test are accompanied by *).

We find that the levels of sodium goes up for both those living in areas with fewer college educated residents, and in places with more African American residents (though such swings were possible before the lockdowns). 
It is known that African American communities in general are consuming more salt than the rest of the American society \cite{james2004factors}, and as African Americans were 30\% more likely to die from heart disease than non-Hispanic whites \cite{omh2021heart}, an increase in its consumption may have serious negative health consequences.
The levels of energy (calories), fat, and saturated fat goes down for those in areas with lower household incomes.
Surveys have found that during lockdowns people have turned to home cooking \cite{bennett2021impact}, however it is not clear whether it corresponded to a reduction in caloric consumption. 
This finding may point to an interest in foods lower in fat as an attempt to better health management.
In similar vein, the levels of sugar go down for areas higher in African American and Hispanic residents.

\begin{table}[t]
\centering

\begin{tabular}{llrr}
\toprule 
Demographic & Nutrient & b$_{covid}$ & $p_{covid}$ \\
\midrule
$\downarrow$ perc. some college & Sodium & 384 & 3.9e-4 \\
$\uparrow$ perc. African American & Sodium & 347 & 1.6e-5 \\
$\downarrow$ median househ. income & Sat. Fat & -1.07 & *4.4e-9 \\
$\uparrow$ perc. African American & Sugar & -1.11 & 2.9e-3 \\
$\uparrow$ perc. Hispanic & Sugar & -1.89 & *2.6e-8 \\
$\downarrow$ median househ. income & Fat & -2.26 & *4.8e-9 \\
$\downarrow$ median househ. income & Energy & -18.27 & *2.7e-6 \\
\bottomrule
\end{tabular}
\caption{Significant changes in macronutrient value of posting in subsets of users: $\downarrow$ below median, $\uparrow$ above median. P values passing random date test are accompanied by *.}
\label{tab:nutrients_demogs}
\end{table}

\subsection{Interest Shift in Food during COVID-19 Using Video Content}

We further examine whether the keywords mentioned in the videos have different co-occurring patterns that changed due to COVID-19 by leveraging Word2Vec embeddings.
In particular, we focus on the mentions of health-related words around diet and comfort foods. 
Table~\ref{tab:word2vecVideos} lists the top-10 keywords that have the highest cosine similarity with the keyword ``food'' and their corresponding word associations before and during the pandemic. 


\begin{table*}[h]
\centering

\footnotesize
\begin{tabular}{p{7.3cm}cp{7.3cm}}
\toprule
\emph{Before COVID-19} & \emph{Keyword} & \emph{During COVID-19} \\
\midrule 
locations, supermarkets, cafes, places, options, menu, buffet, hotels, menus, convenience & \textbf{restaurants} & \textbf{\emph{takeout}}, cafes, supermarkets, places, hotels, menus, jamaica, ocho, asian, japan \\

poultry, vegetables, hamburger, flesh, fish, animal, picky, dairy, seafood, beef & \textbf{meat} & \textbf{\emph{pasture}}, \textbf{\emph{beyond}}, poultry, eaters, beef, eater, seafood, chickens, animal, dairy \\

salads, veggies, starchy, carrots, greens, cruciferous, steamed, asparagus, leafy, fruits & \textbf{vegetables} & \textbf{\emph{beets}}, greens, fruits, \textbf{\emph{kale}}, starchy, cruciferous, salads, carrots, leafy, veggies \\

tasty, satisfying, nutritionally, nourishing, satiating, wholesome, healthy, versatile, healthful, delicious & \textbf{nutritious} & \textbf{\emph{packed}}, \textbf{\emph{balanced}}, \textbf{\emph{comforting}}, nourishing, satisfying, tasty, \textbf{\emph{hearty}}, versatile, \textbf{\emph{convenient}}, wholesome \\

wellness, foodist, link, delights, maestro, markus, zeroing, virta, foodists, sharree & \textbf{thehealthylife} & \textbf{\emph{instafood}}, \textbf{\emph{veganfood}}, \textbf{\emph{fullyraw}}, \textbf{\emph{easyveganrecipes}}, \textbf{\emph{paleohack}}, foodist, \textbf{\emph{govegan}}, delights, hotforfood, bish \\

overeating, unnecessary, fattening, processed, junk, restrictive, avoiding, unnatural, avoided, inherently & \textbf{unhealthy} & \textbf{\emph{crave}}, consuming, \textbf{\emph{moderation}}, restrictive, \textbf{\emph{nutritionally}}, likely, processed, foodstuffs, \textbf{\emph{overeat}}, avoiding \\

mangoes, fruits, tropical, fresh, seedless, melons, specialty, citrus, winter, harvested & \textbf{seasonal} & \textbf{\emph{pomegranates}}, citrus, \textbf{\emph{pears}}, \textbf{\emph{cruciferous}}, \textbf{\emph{turnips}}, \textbf{\emph{exotic}}, \textbf{\emph{figs}}, frugivores, beware, tropical \\

unhealthy, cautious, restrict, focusing, regularly, minimize, intuitive, limiting, intuitively, restricting & \textbf{mindful} & \textbf{\emph{disorder}}, intuitive, knowing, regardless, \textbf{\emph{emotionally}}, \textbf{\emph{habits}}, intuitively, \textbf{\emph{binge}}, \textbf{\emph{emotional}}, focusing \\

local, supermarket, thrift, specialty, restaurants, stores, shops, supermarkets, shelves, grocery & \textbf{convenience} & supermarket, specialty, shopping, shops, items, \textbf{\emph{bulk}}, stores, grocery, \textbf{\emph{available}}, supermarkets \\

influencers, transitioning, veganism, youtubers, activism, trends, interviews, events, tutorials, transitioned & \textbf{challenges} & \textbf{\emph{relationships}}, \textbf{\emph{education}}, \textbf{\emph{opportunities}}, \textbf{\emph{overcome}}, trends, \textbf{\emph{experiences}}, awareness, events, aspects, programs \\

\bottomrule
\end{tabular}
\caption{Word associations obtained through Word2Vec embeddings built using the metadata attached with the videos for example, title, description, and transcript; the bold italicized keywords during COVID-19 emphasize the shift of vocabulary and their relevancy to the pandemic.}
\label{tab:word2vecVideos}
\end{table*}

Some of the key observations made from Table~\ref{tab:word2vecVideos} are: 

\begin{itemize}
\item When we examine the associations with the keyword ``restaurants'', words such as ``buffet'' have been replaced with ``takeout''. This is logical as to follow proper social distancing protocols, restaurants were allowed to have only takeout as an option. 
\item Considering keyword ``meat'', the table highlights the associations such as ``pasture'', ``beyond'', ``chickens'', etc. Studies~\cite{karsten2010vitamins} show that pasture-raised hens produce healthier eggs that contain higher levels of Omega-3 fat, vitamins D and E, and betacarotene. On the other hand, Beyond meat is a producer of plant-based meat substitutes. These word associations may suggest that individuals may have shifted interest to healthier meat/poultry during COVID-19. 
\item When we consider the term ``vegetables'' the word associations are mostly similar before and during COVID-19 except healthier vegetables such as ``beets'' and ``kale''. 

\item  We purposefully didn't perform stemming on the words because we want to see if there are terms that highlight their tense. It is interesting to see word associations with ``nutritious'' during COVID-19 such as ``comforting'', ``hearty'', and ``convenient''. These associations may be due the increased need of emotional calming due to the increased anxiety during that period. 

\item When considering associations with ``thehealthylife'', we find many more diet-related keywords during COVID-19, including ``easyveganrecipes'', ``paleohack'', ``govegan'' instead of influencer-centric keywords before, such as ``foodist'', ``link'', and ``sharree'', further pointing to an increased demand for diet-related content. It is also possible that the lockdowns affected the ability of influencers to produce new content. 

\item Considering ``seasonal'' keywords, the pre-COVID period mentioned ``mangoes'', ``melons'' and ``citrus'' (rich in vitamin A, fiber, vitamin C respectively), during COVID-19 we find a greater variety of fruits and vegetables, including  ``pomegranates'', ``turnips'', ``pears'', and ``figs'' (antioxidant, anti-inflammatory and vitamin C rich foods). This may suggest that there is a high interest in learning about different fruits and vegetables and possibly consuming foods that can boost immunity. 

\item The word associations for the term ``mindful'' calls attention to be mindful of not just minimize, limit, restrict or be cautious of unhealthy foods, but also be mindful of ``emotional'' health, ``binge'' eating, and ``habits'' which were very important especially with staying in restricted spaces for a longer periods of time. 

\item The word associations to the term ``challenges'' during COVID-19 highlight some of the difficulties our society has been dealing with including -- \emph{education} (remote learning and the corresponding issues), \emph{relationships} (pandemic has produced high and enduring levels of psychosocial stress for individuals and families across the world~\cite{liu2020psychosocial}), and \emph{opportunities} (there has been a rise of unemployment according to United States Bureau of Labor Statistics\footnote{\url{https://www.bls.gov/opub/mlr/2021/article/unemployment-rises-in-2020-as-the-country-battles-the-covid-19-pandemic.htm}}). It is notable that we find these possibly non-food-related issues in the data ostensibly about food and nutrition, suggesting their salience in that period. 
\end{itemize}

\begin{table*}[h]
\centering

\footnotesize
\begin{tabular}{rlrrlr|rlrrlr}
\toprule
\multicolumn{6}{c}{\emph{Before COVID-19}} & \multicolumn{6}{c}{\emph{During COVID-19}} \\
& word & OR &  & word & OR &  & word & OR &  & word & OR \\
\midrule
1 & challenge & 0.10 & 11 & whole & 0.26 & 1 & glutenfree & 13.77 & 11 & savory & 3.59 \\
2 & gain & 0.12 & 12 & drink & 0.27 & 2 & coronavirus & 9.91 & 12 & cooker & 3.57 \\
3 & cleanse & 0.20 & 13 & pasta & 0.27 & 3 & bombs & 8.82 & 13 & brown & 3.41 \\
4 & vegans & 0.20 & 14 & snacks & 0.28 & 4 & flour & 8.34 & 14 & treat & 3.41 \\
5 & carnivore & 0.21 & 15 & fast & 0.29 & 5 & waffle & 6.52 & 15 & curry & 3.16 \\
6 & test & 0.21 & 16 & taste & 0.29 & 6 & went & 5.31 & 16 & fryer & 3.03 \\
7 & bosh & 0.22 & 17 & trying & 0.29 & 7 & strawberry & 4.59 & 17 & watching & 3.03 \\
8 & high & 0.24 & 18 & diets & 0.30 & 8 & alkaline & 4.25 & 18 & mini & 2.95 \\
9 & ramen & 0.25 & 19 & sandwich & 0.30 & 9 & fight & 4.17 & 19 & salmon & 2.89 \\
10 & vegan & 0.26 & 20 & thin & 0.31 & 10 & mom & 4.00 & 20 & dough & 2.84 \\
\bottomrule
\end{tabular}
\caption{20 words more likely to be used in tweets before COVID-19 (left) and during COVID-19 (right) by odds ratio.}
\label{tab:odd_ratio_covid}
\end{table*}

\begin{table*}[h]
\centering

\footnotesize
\begin{tabular}{rlrrlr|rlrrlr}
\toprule
\multicolumn{6}{c}{Lower income areas} & \multicolumn{6}{c}{Higher income areas} \\
& word & OR &  & word & OR &  & word & OR &  &  word & OR  \\
\midrule
1 & foods & 2.92 & 11 & soup & 1.96 & 1 & best & 0.30 & 11 & eating & 0.62 \\
2 & breakfast & 2.86 & 12 & keto & 1.90 & 2 & check & 0.33 & 12 & cheese & 0.62 \\
3 & curry & 2.82 & 13 & low & 1.88 & 3 & try & 0.41 & 13 & easy & 0.66 \\
4 & egg & 2.29 & 14 & fat & 1.87 & 4 & quarantine & 0.46 & 14 & chocolate & 0.66 \\
5 & bread & 2.25 & 15 & health & 1.87 & 5 & exercise & 0.46 & 15 & cake & 0.68 \\
6 & dinner & 2.14 & 16 & calories & 1.70 & 6 & episode & 0.58 & 16 & ate & 0.68 \\
7 & banana & 2.14 & 17 & meal & 1.65 & 7 & fasting & 0.58 & 17 & video & 0.69 \\
8 & simple & 2.08 & 18 & cook & 1.59 & 8 & vegan & 0.60 & 18 & eat & 0.72 \\
9 & natural & 2.08 & 19 & salad & 1.56 & 9 & youtube & 0.62 & 19 & potato & 0.72 \\
10 & bacon & 1.96 & 20 & carb & 1.54 & 10 & watch & 0.62 & 20 & full & 0.72 \\
\bottomrule
\end{tabular}
\caption{20 words in tweets more likely to be used by those living in less wealthy (left) and more wealthy (right) areas during COVID-19 by odds ratio.}
\label{tab:odd_ratio_wealth}
\end{table*}

The shift in the word associations suggest both the rise of interest in a broader range of healthy foods and diets associated with health (vegan, keto, paleo, etc.), as well as the emphasis on emotional comfort, emotional eating, and ongoing social challenges. 
These word associations also highlight the shift of emphasis from food influencers and their lifestyle to focusing more on enduring the challenges of life during the pandemic through variety of foods, specifically foods that are healthy and comforting.

\subsection{Interest Shift in Food during COVID-19 Using Tweet Content}

We use the tweet content around the video links to assess the change in the \emph{context} around food sharing between the pre- and pandemic periods, again focusing specifically on the trends around diet and comfort food. We first employ odds ratios on the standardized text of the tweets. Table~\ref{tab:odd_ratio_covid} shows the top-20 words used in tweets posted before and during COVID-19, by odds ratio, which are shown along with the word.
Not surprisingly, by far the most distinguishing word during COVID-19 is \emph{coronavirus}, pointing to the fact that those posting the YouTube links do so within the context of the ongoing epidemic. An increased attention to online \emph{reviews} centers around both recipes (\emph{waffle}, \emph{curry}) and home products (\emph{cooker}, \emph{fryer}), with posts like ``Top-Rated 10 Best Air Fryer Reviews 2020'' directing people to lists of products they may be interested in while cooking at home. Further tips about staying healthy at \emph{home} are very popular, with tweets such as ``Eating Keto While Stuck at Home, How to Stop Snacking All Day!'' offering advice on controlling food intake while under lockdown. On the other hand, we can see the words which are less likely to occur during the COVID-19 period in the first two columns of the table: \emph{cleanse}, \emph{vegan}, \emph{diets}, \emph{fast}, etc. Strict nutritional regimes, such as \emph{ketogenic} and \emph{vegan}, are also more likely to be posted before, but not during COVID-19. In contrast to the findings in the previous section, we find that when sharing these videos, Twitter users do not emphasize the diet aspect, as much as individual foods and tools.

Next, we turn our attention to the differences in such word mentions only during COVID-19 between users living in locations having the median household income below or above the median for our dataset (\$60,270). Table \ref{tab:odd_ratio_wealth} shows the top 20 words used more often by users living in less (left) and more (right) wealthy parts of the U.S., by odds ratio. The lower income areas favor the \emph{simple} recipes, such as \emph{curry}, \emph{eggs} and \emph{bread}, and are concerned about the food being \emph{natural} and recipes \emph{low} in \emph{fat}. In higher income areas, people actually mention \emph{quarantine}, and are more likely to mention \emph{best} and \emph{exercise}, as well as \emph{cake} and \emph{chocolate}. We find that the majority of foods mentioned in the low income areas are potentially healthy, and refer to meal times and cooking, indicating a focus on everyday, useful content. Instead, we find mentions of potentially dangerous practices like fasting and comfort foods like chocolate and cake mentioned in higher income areas. 
Note that, when we consider both sets of users, the most frequent terms used by both are very similar: \emph{vegan}, \emph{weight}, \emph{diet}, \emph{keto} and \emph{loss}, pointing to a common interest in weight management strategies across the board. 

The odds-ratio analysis highlights that during COVID-19, high calorie foods are more likely to be used in tweets than before the pandemic. Thus, we investigate further the change in the vocabulary to assess if the word associations changed with food-related keywords. 
Table~\ref{tab:word2vecTweets} shows the words that are semantically and syntactically related to each keyword from the corpus of tweets, retrieved similarly from videos in Table~\ref{tab:word2vecVideos}. 

\begin{itemize}
\item Among the top-10 highly associated words with the keyword ``foodies'' and ``healthyeating'' (as well as other keywords), during pandemic we find numerous references to the social distancing measures such as ``stayathomesavelives'', ``cookingathome'', ``stayhome'', ``healthyathome''. Instead, before COVID-19 we find a larger focus on influencer-specific keywords including ``instafood'', ``instablog'', ``foodblog'', ``foodphotography'', and ``foodblogger'', which decrease during the pandemic. 
\item There is a specific emphasis on cooking (using keywords ``cook'' and ``cooking'') ``easy recipes'' or ``recipe of the day'' that would be helpful for many individuals who might have not made meals at home before the pandemic. Other kinds of emphasis is also on homemade (``homemade'') delicious (``delicious'', ``yummy'', ``tasty'') food. 
\item When we look at keyword ``restaurant'' before the pandemic it was mostly different kinds of restaurants, menu items, details of the service provided, etc. However, during the pandemic the top associated words include ``masked'' and ``delivery''. 
\item The term ``vegetarian'' before the pandemic is focused on terms related to being a ``vegan''. However, during the pandemic the term is mainly associated with specific dishes and ingredients, such as ``soup'', ``lentil'', ``chickpea'', ``burrito'', ``salad'', etc. Echoing our findings from previous section, these keywords suggest an emphasis on easy-to-make, comfort foods instead of generic references to vegetarianism prevalent before COVID-19. 
\end{itemize} 


\begin{table*}[h]
\centering

\footnotesize
\begin{tabular}{p{7.1cm}cp{7.1cm}}
\toprule
\emph{Before COVID-19} & \emph{Keyword} & \emph{During COVID-19} \\
\midrule
foodblogger, homecooking, foodblog, foodphotography, foodlover, instagood, indianfood, nonalcholic, easyrecipes, instafood & \textbf{foodies} & \textbf{\emph{cookingathome}}, foodblog, cookingmaster, \textbf{\emph{stayhome}}, foodporn, \textbf{\emph{easyrecipe}}, \textbf{\emph{healthyathome}}, recipeblog, \textbf{\emph{stayathomesavelives}}, \textbf{\emph{recipesforthepeople}} \\ 

healthyfood, healthyrecipes, cleaneating, eatclean, quickmeals, healthylifestyle, healthtips, healthycooking, healthyliving, quickrecipes & \textbf{healthyeating} & healthyfood, \textbf{\emph{healthyathome}}, healthyliving, \textbf{\emph{stayhome}}, foodies, \textbf{\emph{cookingathome}}, \textbf{\emph{healthymeals}}, cookingmaster, recipesforthepeople, \textbf{\emph{stayathomesavelives}} \\

party, tasty, indian, supportive, dinner, spicy, summer, potatoes, snack, asmr & \textbf{cook} & \textbf{\emph{recipeoftheday}}, yummy, yummyfood, \textbf{\emph{easyrecipes}}, italianfood, \textbf{\emph{homemade}}, foodie, \textbf{\emph{easyrecipe}}, tasty, delicious \\

beer, eats, unbelievable, service, nuggets, kfc, popeyes, options, cafe, greggs & \textbf{restaurant} & pickle, artichokes,  \textbf{\emph{broth}},  \textbf{\emph{mangoes}}, drumstick, bakery, \textbf{\emph{delivery}}, \textbf{\emph{spiced}}, \textbf{\emph{vietnamese}}, \textbf{\emph{masked}} \\

randirobics, veganyoutuber, dareme, veganrecipe, veganfood, oliver, cheflife, rawvegan, recipeoftheday, options & \textbf{vegetarian} & lentil, \textbf{\emph{soup}}, chickpea, \textbf{\emph{burrito}}, \textbf{\emph{pasta}}, \textbf{\emph{salad}}, \textbf{\emph{chickpeas}}, rice, dinner, creamy \\

cheflife, foodies, indianfood, recipeoftheday, party, tasty, lunchbox, foodporn, foodtruck, foodie & \textbf{cooking} & foodie, recipeoftheday, yummy, foodporn, foodies, yummyfood, \textbf{\emph{cookwithme}}, \textbf{\emph{easyrecipes}}, \textbf{\emph{cleaneating}}, easyrecipe \\
 
recipeblog, easyrecipes, cheflife, recipevideo, yummy, foodblog, foodphotography, foodie, foodpics, foodblogger & \textbf{recipeoftheday} & cook, yummy, yummyfood, easyrecipes, foodie, \textbf{\emph{cookwithme}}, italianfood, easyrecipe, \textbf{\emph{quarantinecooking}}, cooking \\

youtubechannel, quickmeals, writingcommunity, newvideo, vlogger, youtubetrends, whatieatinaday, youtubevideos, foodporn, blogger & \textbf{youtubers} & \textbf{\emph{ramayanonddnational}}, \textbf{\emph{mondayvibes}}, \textbf{\emph{saturdaymotivation}}, saturdaykitchen, \textbf{\emph{thursdaythoughts}}, \textbf{\emph{ramayan}}, \textbf{\emph{lockdownextension}}, poha, howto, \textbf{\emph{thursdaymotivation}} \\

unbelievable, bakes, street, options, chinese, party, supportive, lovers, pantry, lunchbox & \textbf{comfort} & singapore, \textbf{\emph{takeaway}}, \textbf{\emph{soups}}, \textbf{\emph{snack}}, bar, \textbf{\emph{chilli}}, caribbean, flowers, papaya, oatmeal \\

vegetables, fruits, prepare, friendly, fermented, cleanest, fridge, dishes, lunchbox, mediterranean & \textbf{veggies} & substitutes, using, sticks, side, cafe, unique, \textbf{\emph{nuts}}, \textbf{\emph{pea}}, \textbf{\emph{celery}}, choices \\
\bottomrule
\end{tabular}
\caption{Top-10 words associated with each keyword obtained using Word2Vec embeddings generated from the tweet text. The bold italicized keywords during COVID-19 emphasize examples in the shift of the vocabulary.}
\label{tab:word2vecTweets}
\end{table*}

\section{Discussion and Conclusion}

Overall, we find mixed results concerning dieting and comfort foods.
On one hand, we find encouraging signs throughout the study: the fact that overall newly mentioned foods have lower nutritional values of energy, fat, and saturated fat, and a focus on health remains, including in lower income areas.
This finding may be linked to a possible decrease in fast food consumption, which has been documented outside the U.S.~\cite{ruiz2020covid}.
On the other, we find the levels of sodium increase for both those living in areas with fewer college educated residents and in places with more African American residents. 
This is a concerning trend, as over-consumption of sodium may lead to high blood pressure, heart disease, and stroke\footnote{\url{https://tinyurl.com/22jyfueb}}, ailments already disproportionally afflicting the African American community.
Although we cannot equate social media sharing with actual consumption, previous studies have indicated a moderate to strong relationship between posted foods and local health outcomes~\cite{min2019survey}. 
The heterogeneity of our findings point to the necessity of further contextualization, possibly by the use of surveys and interviews, in other demographics which may affect the severity of lockdown impact, such as gender~\cite{ozenouglu2021changes}, employment status, and psychological state~\cite{paltrinieri2021beyond}. 

The analysis of words associated with terms related to food and nutrition revealed a shift in emphasis from generic diets and lifestyle content to concrete examples of foods, recipes, and activities that constitute the daily nutritional habits of those under lockdown. 
In particular, we found that before the pandemic the emphasis was mainly on different kinds of diets such as vegan diet, keto diet, rather than specific ingredients that constitute such a diet, influencer-specific details such as their lifestyle, challenges, vlogging, etc, and high-level particulars of food industry such as restaurant locations and types of restaurants. 
However, during the pandemic, focus has been shifted to numerous references to the specific ingredients, and emphasis on easy-to-make, comfort foods.
When we consider the video-based Word2Vec associations to the food-related terms during COVID-19, we find words that emphasize healthy and nutritious foods. More specifically, we found that shift in the word associations suggest both the increase in interest in a broad range of foods and diets as well as emphasis on emotional comfort, emotional eating and other pandemic-related social challenges.
Instead, when we analyzed the tweet content, odds ratio analysis showed that calorie rich foods (e.g. comfort food that may not be healthy in large quantities) become more popular during COVID-19 compared to before.
This suggests that, while healthy foods remain popular during the pandemic, comfort food mentions became relatively more popular.
It would be interesting to see if this trend continues during subsequent lockdowns.

Outside the nutrition information, content analysis of the tweets revealed an array of coping mechanisms around the isolation during lockdowns, including offering online support to other users, sharing easy recipes, and sharing takeaway food experiences, highlighting the community togetherness during the COVID-19 pandemic.
Indeed, numerous organizations have provided online resources, such as directions to donate to the local food pantry by Save the Children\footnote{\url{https://tinyurl.com/55hsc45a}}, mental health resources by CDC\footnote{\url{https://tinyurl.com/yct62tkn}} and UK's NHS\footnote{\url{https://tinyurl.com/3xhft8pc}}, and various community grants\footnote{\url{https://tinyurl.com/yppmkdat}}.
Although it was not the focus of this research, the community engagement was a surprising finding, and we would encourage future researchers to further examine the resources, and especially potential calls for help, which may be present on social media. 
Alternatively, it would be interesting to perform a quality control on the videos of this dataset, as diet-related COVID-19 ``cures'' and related misinformation are known to have circulated at around this time~\cite{el2020let}.

This study illustrates how the combination of information around food-focused tweets and videos to which they connect can highlight the dietary interest shifts over time. 
Tools presented here may supplement the existing health monitoring infrastructure used by public health organizations, including surveys and interviews. 
Open-ended nature of such observational studies may reveal trends which are not yet on the radar (such as new trendy diets), as well as the social context around the needs and concerns of those affected by public health interventions (e.g. need for simple recipes for novice cooks). Timely and effective responses to these needs will help alleviate potential negative impacts of the unprecedented COVID-19 interventions.

The reader should note that the data examined in this study is biased in multiple ways, potentially constraining generalizability of these findings. 
Social media users are likely to be younger and more tech-savvy, and especially those actively posting (instead of passively browsing)~\cite{mellon2017twitter,Mustafaraj2011}.
The peculiar timing of the COVID-19 onset may have coincided with other seasonal nutritional behaviors (such as a possible increase in the availability of fresh fruit and vegetables).
A comparison with similar times in other years may provide an insight as to the effect of such peculiarities.
Further, the fact that the users captured in this dataset used two social media potentially puts them in a ``vocal minority'' of vocal influencers~\cite{Mustafaraj2011}.
Alternative data sources on the viewership and browsing activity of other users would be invaluable to gauge the interest of others less inclined to post content (such as was done recently in the context of political content \cite{hosseinmardi2021examining}).
In future work it would be interesting to compare the data to similar time in other years in order to control for potential seasonal confounders.

\section{Broader Impact}

The broader aim of this work is to contribute to the understanding of health behaviors using user generated content and computational tools, specifically, the insights and tools we present here apply to a broad spectrum of stakeholders. 
First and foremost, the research design of this study attempts to minimize any harm to the users whose content was captured in this research, as their usernames and other identifiable information was excluded from the report and the analysis.
Still, various high-risk groups could be captured here, such as minors (despite the platforms trying to enforce age restrictions), those struggling with eating disorders, or lacking access to healthy and affordable food. 
Secondly, there is a large portion of users who may have been affected by this content who have seen or interacted with it, without being captured by our data collection process. 
Those not captured would also include those with disabilities, and who were unable to use these platforms due to lack of access. 
Furthermore, findings here may impact the decision-making of public health researchers, both in terms of insights into information sharing behavior during COVID-19, and in terms of the scope of the available technology for monitoring such behavior.
As such, we are aware that tools for public social media data surveillance may be mis-used in order to target people or engage in other kinds of surveillance -- note that such behavior would be against the Terms of Service of both Twitter \cite{twitterpolicy}  
and YouTube \cite{YouTubepolicy}. 
Still, in an effort to provide transparency in the research, we will make the dataset available in accordance with the above Terms of Service, and along with the manual and automatic annotations necessary for the analysis.

\section{Acknowledgments}
The authors acknowledge support from the Lagrange Project of the Institute for Scientific Interchange Foundation (ISI Foundation) funded by Fondazione Cassa di Risparmio di Torino (Fondazione CRT), and startup funds through the Rensselaer-IBM AIRC.



{
\fontsize{9.1pt}{10.8pt} \selectfont
\bibliography{dietube}

\begin{thebibliography}{58}
\providecommand{\natexlab}[1]{#1}

\bibitem[{Abbar, Mejova, and Weber(2015)}]{abbar2015you}
Abbar, S.; Mejova, Y.; and Weber, I. 2015.
\newblock You tweet what you eat: Studying food consumption through twitter.
\newblock In \emph{Proceedings of the 33rd Annual ACM Conference on Human
  Factors in Computing Systems}, 3197--3206.

\bibitem[{Abisheva et~al.(2014)Abisheva, Garimella, Garcia, and
  Weber}]{abisheva2014watches}
Abisheva, A.; Garimella, V. R.~K.; Garcia, D.; and Weber, I. 2014.
\newblock Who watches (and shares) what on youtube? and when? using twitter to
  understand youtube viewership.
\newblock In \emph{Proceedings of the 7th ACM international conference on Web
  search and data mining}, 593--602.

\bibitem[{Alajajian et~al.(2017)Alajajian, Williams, Reagan, Alajajian, Frank,
  Mitchell, Lahne, Danforth, and Dodds}]{alajajian2017lexicocalorimeter}
Alajajian, S.~E.; Williams, J.~R.; Reagan, A.~J.; Alajajian, S.~C.; Frank,
  M.~R.; Mitchell, L.; Lahne, J.; Danforth, C.~M.; and Dodds, P.~S. 2017.
\newblock The Lexicocalorimeter: Gauging public health through caloric input
  and output on social media.
\newblock \emph{PloS one}, 12(2): e0168893.

\bibitem[{Bennett et~al.(2021)Bennett, Young, Butler, and
  Coe}]{bennett2021impact}
Bennett, G.; Young, E.; Butler, I.; and Coe, S. 2021.
\newblock The impact of lockdown during the COVID-19 outbreak on dietary habits
  in various population groups: a scoping review.
\newblock \emph{Frontiers in nutrition}, 8: 53.

\bibitem[{Bernal, Cummins, and Gasparrini(2017)}]{bernal2017interrupted}
Bernal, J.~L.; Cummins, S.; and Gasparrini, A. 2017.
\newblock Interrupted time series regression for the evaluation of public
  health interventions: a tutorial.
\newblock \emph{International journal of epidemiology}, 46(1): 348--355.

\bibitem[{{Christodoulou, George and Georgiou, Chryssis and Pallis,
  George}(2012)}]{christodoulou2012youtubediffusiontwitter}
{Christodoulou, George and Georgiou, Chryssis and Pallis, George}. 2012.
\newblock The Role of Twitter in YouTube Videos Diffusion.
\newblock In {Wang, X. Sean and Cruz, Isabel and Delis, Alex and Huang,
  Guangyan}, ed., \emph{{Web Information Systems Engineering - (WISE)}},
  426--439.

\bibitem[{Christofaro et~al.(2021)Christofaro, Werneck, Tebar, Lofrano-Prado,
  Botero, Cucato, Malik, Correia, Ritti-Dias, and
  Prado}]{christofaro2021physical}
Christofaro, D.~G.; Werneck, A.~O.; Tebar, W.~R.; Lofrano-Prado, M.~C.; Botero,
  J.~P.; Cucato, G.~G.; Malik, N.; Correia, M.~A.; Ritti-Dias, R.~M.; and
  Prado, W.~L. 2021.
\newblock Physical activity is associated with improved eating habits during
  the COVID-19 pandemic.
\newblock \emph{Frontiers in Psychology}, 12.

\bibitem[{Chung et~al.(2017)Chung, Agapie, Schroeder, Mishra, Fogarty, and
  Munson}]{chung2017personal}
Chung, C.-F.; Agapie, E.; Schroeder, J.; Mishra, S.; Fogarty, J.; and Munson,
  S.~A. 2017.
\newblock When personal tracking becomes social: Examining the use of Instagram
  for healthy eating.
\newblock In \emph{Proceedings of the 2017 CHI Conference on human factors in
  computing systems}, 1674--1687.

\bibitem[{De~Choudhury, Sharma, and Kiciman(2016)}]{de2016characterizing}
De~Choudhury, M.; Sharma, S.; and Kiciman, E. 2016.
\newblock Characterizing dietary choices, nutrition, and language in food
  deserts via social media.
\newblock In \emph{Proceedings of the 19th acm conference on computer-supported
  cooperative work \& social computing}, 1157--1170.

\bibitem[{Devlin et~al.(2018)Devlin, Chang, Lee, and
  Toutanova}]{devlin2018bert}
Devlin, J.; Chang, M.-W.; Lee, K.; and Toutanova, K. 2018.
\newblock Bert: Pre-training of deep bidirectional transformers for language
  understanding.
\newblock \emph{arXiv preprint arXiv:1810.04805}.

\bibitem[{Di~Renzo et~al.(2020)Di~Renzo, Gualtieri, Pivari, Soldati,
  Attin{\`a}, Cinelli, Leggeri, Caparello, Barrea, Scerbo
  et~al.}]{di2020eating}
Di~Renzo, L.; Gualtieri, P.; Pivari, F.; Soldati, L.; Attin{\`a}, A.; Cinelli,
  G.; Leggeri, C.; Caparello, G.; Barrea, L.; Scerbo, F.; et~al. 2020.
\newblock Eating habits and lifestyle changes during COVID-19 lockdown: an
  Italian survey.
\newblock \emph{Journal of Translational Medicine}, 18(1): 1--15.

\bibitem[{{Economic Research Service. US Department of
  Agriculture}(2020)}]{usda2020food}
{Economic Research Service. US Department of Agriculture}. 2020.
\newblock Food Security Status of U.S. Households in 2020.
\newblock Available from:
  \url{https://www.ers.usda.gov/topics/food-nutrition-assistance/food-security-in-the-us/key-statistics-graphics.aspx}.
\newblock (accessed Jan 6, 2022).

\bibitem[{El~Ghoch, Valerio et~al.(2020)}]{el2020let}
El~Ghoch, M.; Valerio, A.; et~al. 2020.
\newblock Let food be the medicine, but not for coronavirus: Nutrition and food
  science, telling myths from facts.
\newblock \emph{Journal of population therapeutics and clinical pharmacology},
  27(SP1): e1--e4.

\bibitem[{Esobi, Lasode, and Barriguete(2020)}]{esobi2020impact}
Esobi, I.~C.; Lasode, M.; and Barriguete, M.~F. 2020.
\newblock The Impact of COVID-19 on Healthy Eating Habits.
\newblock \emph{J Clin Nutr Heal}, 1(1): 001--002.

\bibitem[{Fowles et~al.(2012)Fowles, Stang, Bryant, and Kim}]{fowles2012stress}
Fowles, E.~R.; Stang, J.; Bryant, M.; and Kim, S. 2012.
\newblock Stress, depression, social support, and eating habits reduce diet
  quality in the first trimester in low-income women: a pilot study.
\newblock \emph{Journal of the Academy of Nutrition and Dietetics}, 112(10):
  1619--1625.

\bibitem[{Fried et~al.(2014)Fried, Surdeanu, Kobourov, Hingle, and
  Bell}]{fried2014analyzing}
Fried, D.; Surdeanu, M.; Kobourov, S.; Hingle, M.; and Bell, D. 2014.
\newblock Analyzing the language of food on social media.
\newblock In \emph{2014 IEEE International Conference on Big Data (Big Data)},
  778--783.

\bibitem[{Gore, Diallo, and Padilla(2015)}]{gore2015you}
Gore, R.~J.; Diallo, S.; and Padilla, J. 2015.
\newblock You are what you tweet: connecting the geographic variation in
  america’s obesity rate to Twitter content.
\newblock \emph{PloS one}, 10(9): e0133505.

\bibitem[{Hingle et~al.(2013)Hingle, Yoon, Fowler, Kobourov, Schneider, Falk,
  and Burd}]{hingle2013collection}
Hingle, M.; Yoon, D.; Fowler, J.; Kobourov, S.; Schneider, M.~L.; Falk, D.; and
  Burd, R. 2013.
\newblock Collection and visualization of dietary behavior and reasons for
  eating using Twitter.
\newblock \emph{Journal of medical Internet research}, 15(6): e125.

\bibitem[{Ho, Paultre, and Mosca(2002)}]{ho2002lifestyle}
Ho, J.~E.; Paultre, F.; and Mosca, L. 2002.
\newblock Lifestyle changes in New Yorkers after September 11, 2001 (data from
  the post-disaster Heart Attack Prevention Program).
\newblock \emph{American Journal of Cardiology}, 90(6): 680--682.

\bibitem[{Hosseinmardi et~al.(2021)Hosseinmardi, Ghasemian, Clauset, Mobius,
  Rothschild, and Watts}]{hosseinmardi2021examining}
Hosseinmardi, H.; Ghasemian, A.; Clauset, A.; Mobius, M.; Rothschild, D.~M.;
  and Watts, D.~J. 2021.
\newblock Examining the consumption of radical content on YouTube.
\newblock \emph{Proceedings of the National Academy of Sciences}, 118(32).

\bibitem[{Israel and Briones(2012)}]{israel2012impacts}
Israel, D.~C.; and Briones, R.~M. 2012.
\newblock Impacts of natural disasters on agriculture, food security, and
  natural resources and environment in the Philippines.
\newblock Technical report, PIDS discussion paper series.

\bibitem[{Kar et~al.(2021)Kar, Motoyama, Carrel, Miller, and Le}]{kar2021covid}
Kar, A.; Motoyama, Y.; Carrel, A.~L.; Miller, H.~J.; and Le, H.~T. 2021.
\newblock COVID-19 exacerbates unequal food access.
\newblock \emph{Applied Geography}, 134: 102517.

\bibitem[{Karami et~al.(2018)Karami, Dahl, Turner-McGrievy, Kharrazi, and
  Shaw~Jr}]{karami2018characterizing}
Karami, A.; Dahl, A.~A.; Turner-McGrievy, G.; Kharrazi, H.; and Shaw~Jr, G.
  2018.
\newblock Characterizing diabetes, diet, exercise, and obesity comments on
  Twitter.
\newblock \emph{International Journal of Information Management}, 38(1): 1--6.

\bibitem[{Karsten et~al.(2010)Karsten, Patterson, Stout, and
  Crews}]{karsten2010vitamins}
Karsten, H.; Patterson, P.; Stout, R.; and Crews, G. 2010.
\newblock Vitamins A, E and fatty acid composition of the eggs of caged hens
  and pastured hens.
\newblock \emph{Renewable Agriculture and Food Systems}, 25(1): 45--54.

\bibitem[{Kenter et~al.(2015)Kenter, Wevers, Huijnen, and
  De~Rijke}]{kenter2015ad}
Kenter, T.; Wevers, M.; Huijnen, P.; and De~Rijke, M. 2015.
\newblock Ad hoc monitoring of vocabulary shifts over time.
\newblock In \emph{Proceedings of the 24th ACM international on conference on
  information and knowledge management}, 1191--1200.

\bibitem[{Kwak et~al.(2010)Kwak, Lee, Park, and Moon}]{kwak2010twitter}
Kwak, H.; Lee, C.; Park, H.; and Moon, S. 2010.
\newblock What is Twitter, a social network or a news media?
\newblock In \emph{Proceedings of the 19th international conference on World
  wide web}, 591--600.

\bibitem[{Lee et~al.(2019)Lee, Achananuparp, Liu, Lim, and
  Varshney}]{lee2019estimating}
Lee, H.~H.; Achananuparp, P.; Liu, Y.; Lim, E.-P.; and Varshney, L.~R. 2019.
\newblock Estimating glycemic impact of cooking recipes via online
  crowdsourcing and machine learning.
\newblock In \emph{Proceedings of the 9th International Conference on Digital
  Public Health}, 31--35.

\bibitem[{Liu and Doan(2020)}]{liu2020psychosocial}
Liu, C.~H.; and Doan, S.~N. 2020.
\newblock Psychosocial stress contagion in children and families during the
  COVID-19 pandemic.
\newblock \emph{Clinical pediatrics}, 59(9-10): 853--855.

\bibitem[{Liu, Medlar, and Glowacka(2021)}]{liu2021statistically}
Liu, Y.; Medlar, A.; and Glowacka, D. 2021.
\newblock Statistically significant detection of semantic shifts using
  contextual word embeddings.
\newblock \emph{arXiv preprint arXiv:2104.03776}.

\bibitem[{Manikonda, Meduri, and Kambhampati(2016)}]{manikonda2016tweeting}
Manikonda, L.; Meduri, V.~V.; and Kambhampati, S. 2016.
\newblock Tweeting the mind and instagramming the heart: Exploring
  differentiated content sharing on social media.
\newblock In \emph{Tenth international AAAI conference on web and social
  media}.

\bibitem[{Mattioli et~al.(2020)Mattioli, Puviani, Nasi, and
  Farinetti}]{mattioli2020covid}
Mattioli, A.~V.; Puviani, M.~B.; Nasi, M.; and Farinetti, A. 2020.
\newblock COVID-19 pandemic: the effects of quarantine on cardiovascular risk.
\newblock \emph{European journal of clinical nutrition}, 74(6): 852--855.

\bibitem[{McIntosh et~al.(1980)}]{mcintosh1980food}
McIntosh, C.~E.; et~al. 1980.
\newblock Food and nutrition problems associated with natural disasters.
\newblock \emph{Cajanus}, 13(1): 18--27.

\bibitem[{McKenzie, Schargrodsky et~al.(2005)}]{mckenzie2005buying}
McKenzie, D.; Schargrodsky, E.; et~al. 2005.
\newblock Buying less, but shopping more: Changes in consumption patterns
  during a crisis.
\newblock \emph{Bureau for Research and Economic Analysis of Development
  (BREAD) Working Paper}, 92.

\bibitem[{Mejova et~al.(2015)Mejova, Haddadi, Noulas, and
  Weber}]{mejova2015foodporn}
Mejova, Y.; Haddadi, H.; Noulas, A.; and Weber, I. 2015.
\newblock \# foodporn: Obesity patterns in culinary interactions.
\newblock In \emph{Proceedings of the 5th International Conference on Digital
  Health 2015}, 51--58.

\bibitem[{Mejova and Kourtellis(2021)}]{mejova2021youtubing}
Mejova, Y.; and Kourtellis, N. 2021.
\newblock YouTubing at Home: Media Sharing Behavior Change as Proxy for
  MobilityAround COVID-19 Lockdowns.
\newblock \emph{Proceedings of the 2021 ACM on Web Science Conference}.

\bibitem[{Mellon and Prosser(2017)}]{mellon2017twitter}
Mellon, J.; and Prosser, C. 2017.
\newblock Twitter and Facebook are not representative of the general
  population: Political attitudes and demographics of British social media
  users.
\newblock \emph{Research \& Politics}, 4(3).

\bibitem[{Michels et~al.(2012)Michels, Sioen, Braet, Eiben, Hebestreit,
  Huybrechts, Vanaelst, Vyncke, and De~Henauw}]{michels2012stress}
Michels, N.; Sioen, I.; Braet, C.; Eiben, G.; Hebestreit, A.; Huybrechts, I.;
  Vanaelst, B.; Vyncke, K.; and De~Henauw, S. 2012.
\newblock Stress, emotional eating behaviour and dietary patterns in children.
\newblock \emph{Appetite}, 59(3): 762--769.

\bibitem[{Mikolov et~al.(2013)Mikolov, Sutskever, Chen, Corrado, and
  Dean}]{mikolov2013distributed}
Mikolov, T.; Sutskever, I.; Chen, K.; Corrado, G.~S.; and Dean, J. 2013.
\newblock Distributed representations of words and phrases and their
  compositionality.
\newblock In \emph{Advances in neural information processing systems},
  3111--3119.

\bibitem[{Min et~al.(2019)Min, Jiang, Liu, Rui, and Jain}]{min2019survey}
Min, W.; Jiang, S.; Liu, L.; Rui, Y.; and Jain, R. 2019.
\newblock A survey on food computing.
\newblock \emph{ACM Computing Surveys (CSUR)}, 52(5): 1--36.

\bibitem[{Morgan, Lampe, and Shafiq(2013)}]{morgan2013news}
Morgan, J.~S.; Lampe, C.; and Shafiq, M.~Z. 2013.
\newblock Is news sharing on Twitter ideologically biased?
\newblock In \emph{Proceedings of the 2013 conference on Computer supported
  cooperative work}, 887--896.

\bibitem[{Mustafaraj et~al.(2011)Mustafaraj, Finn, Whitlock, and
  Metaxas}]{Mustafaraj2011}
Mustafaraj, E.; Finn, S.; Whitlock, C.; and Metaxas, P.~T. 2011.
\newblock {Vocal Minority versus Silent Majority: Discovering the Opinions of
  the Long Tail}.
\newblock In \emph{International Conference on Social Computing}, 103--110.
  IEEE.

\bibitem[{Nguyen et~al.(2017)Nguyen, Meng, Li, Kath, McCullough, Paul,
  Kanokvimankul, Nguyen, and Li}]{nguyen2017social}
Nguyen, Q.; Meng, H.; Li, D.; Kath, S.; McCullough, M.; Paul, D.;
  Kanokvimankul, P.; Nguyen, T.; and Li, F. 2017.
\newblock Social media indicators of the food environment and state health
  outcomes.
\newblock \emph{Public Health}, 148: 120--128.

\bibitem[{Ofli et~al.(2017)Ofli, Aytar, Weber, Al~Hammouri, and
  Torralba}]{ofli2017saki}
Ofli, F.; Aytar, Y.; Weber, I.; Al~Hammouri, R.; and Torralba, A. 2017.
\newblock Is saki\# delicious? the food perception gap on instagram and its
  relation to health.
\newblock In \emph{Proceedings of the 26th International Conference on World
  Wide Web}, 509--518.

\bibitem[{{\"O}zeno{\u{g}}lu et~al.(2021){\"O}zeno{\u{g}}lu, {\c{C}}evik,
  {\c{C}}olak, Alt{\i}nta{\c{s}}, and Alaku{\c{s}}}]{ozenouglu2021changes}
{\"O}zeno{\u{g}}lu, A.; {\c{C}}evik, E.; {\c{C}}olak, H.; Alt{\i}nta{\c{s}},
  T.; and Alaku{\c{s}}, K. 2021.
\newblock Changes in nutrition and lifestyle habits during the COVID-19
  pandemic in Turkey and the effects of healthy eating attitudes.
\newblock \emph{Mediterranean Journal of Nutrition and Metabolism}, (Preprint):
  1--17.

\bibitem[{Paltrinieri et~al.(2021)Paltrinieri, Bressi, Costi, Mazzini, Cavuto,
  Ottone, De~Panfilis, Fugazzaro, Rondini, and
  Giorgi~Rossi}]{paltrinieri2021beyond}
Paltrinieri, S.; Bressi, B.; Costi, S.; Mazzini, E.; Cavuto, S.; Ottone, M.;
  De~Panfilis, L.; Fugazzaro, S.; Rondini, E.; and Giorgi~Rossi, P. 2021.
\newblock Beyond Lockdown: The Potential Side Effects of the SARS-CoV-2
  Pandemic on Public Health.
\newblock \emph{Nutrients}, 13(5): 1600.

\bibitem[{vrehuuvrek, Sojka et~al.(2011)}]{vrehuuvrek2011gensim}
vrehuuvrek, R.; Sojka, P.; et~al. 2011.
\newblock Gensim—statistical semantics in python.
\newblock Retrieved from \url{genism.org}.

\bibitem[{Rokicki, Trattner, and Herder(2018)}]{rokicki2018impact}
Rokicki, M.; Trattner, C.; and Herder, E. 2018.
\newblock The impact of recipe features, social cues and demographics on
  estimating the healthiness of online recipes.
\newblock In \emph{Twelfth International AAAI Conference on Web and Social
  Media}.

\bibitem[{Ruiz-Roso et~al.(2020)Ruiz-Roso, de~Carvalho~Padilha,
  Mantilla-Escalante, Ulloa, Brun, Acevedo-Correa, Arantes Ferreira~Peres,
  Martorell, Aires, de~Oliveira~Cardoso et~al.}]{ruiz2020covid}
Ruiz-Roso, M.~B.; de~Carvalho~Padilha, P.; Mantilla-Escalante, D.~C.; Ulloa,
  N.; Brun, P.; Acevedo-Correa, D.; Arantes Ferreira~Peres, W.; Martorell, M.;
  Aires, M.~T.; de~Oliveira~Cardoso, L.; et~al. 2020.
\newblock Covid-19 Confinement and Changes of Adolescent’s Dietary Trends in
  Italy, Spain, Chile, Colombia and Brazil.
\newblock \emph{Nutrients}, 12(6): 1807.

\bibitem[{Salvador et~al.(2017)Salvador, Hynes, Aytar, Marin, Ofli, Weber, and
  Torralba}]{salvador2017learning}
Salvador, A.; Hynes, N.; Aytar, Y.; Marin, J.; Ofli, F.; Weber, I.; and
  Torralba, A. 2017.
\newblock Learning cross-modal embeddings for cooking recipes and food images.
\newblock In \emph{Proceedings of the IEEE conference on computer vision and
  pattern recognition}, 3020--3028.

\bibitem[{Teixeira et~al.(2021)Teixeira, Vitorino, da~Silva, Raposo, Aquino,
  and Ribas}]{teixeira2021eating}
Teixeira, M.~T.; Vitorino, R.~S.; da~Silva, J.~H.; Raposo, L.~M.; Aquino, L.
  A.~d.; and Ribas, S.~A. 2021.
\newblock Eating habits of children and adolescents during the COVID-19
  pandemic: The impact of social isolation.
\newblock \emph{Journal of Human Nutrition and Dietetics}.

\bibitem[{Trattner and Elsweiler(2017)}]{trattner2017investigating}
Trattner, C.; and Elsweiler, D. 2017.
\newblock Investigating the healthiness of internet-sourced recipes:
  implications for meal planning and recommender systems.
\newblock In \emph{Proceedings of the 26th international conference on world
  wide web}, 489--498.

\bibitem[{Tryon et~al.(2013)Tryon, Carter, DeCant, and
  Laugero}]{tryon2013chronic}
Tryon, M.~S.; Carter, C.~S.; DeCant, R.; and Laugero, K.~D. 2013.
\newblock Chronic stress exposure may affect the brain's response to high
  calorie food cues and predispose to obesogenic eating habits.
\newblock \emph{Physiology \& behavior}, 120: 233--242.

\bibitem[{Twitter(2022)}]{twitterpolicy}
Twitter. 2022.
\newblock Developer Agreement and Policy.
\newblock
  https://developer.twitter.com/en/developer-terms/agreement-and-policy.
\newblock (accessed Jan 6, 2022).

\bibitem[{{US Department of Health and Human Services}(2021)}]{omh2021heart}
{US Department of Health and Human Services}. 2021.
\newblock Heart Disease and African Americans.
\newblock Available from:
  \url{https://minorityhealth.hhs.gov/omh/browse.aspx?lvl=4&lvlid=19}.
\newblock (accessed Jan 7, 2022).

\bibitem[{Vidal et~al.(2015)Vidal, Ares, Mach{\'\i}n, and
  Jaeger}]{vidal2015using}
Vidal, L.; Ares, G.; Mach{\'\i}n, L.; and Jaeger, S.~R. 2015.
\newblock Using Twitter data for food-related consumer research: A case study
  on “what people say when tweeting about different eating situations”.
\newblock \emph{Food Quality and Preference}, 45: 58--69.

\bibitem[{Vydiswaran et~al.(2020)Vydiswaran, Romero, Zhao, Yu, Gomez-Lopez, Lu,
  Iott, Baylin, Jansen, Clarke et~al.}]{vydiswaran2020uncovering}
Vydiswaran, V.~V.; Romero, D.~M.; Zhao, X.; Yu, D.; Gomez-Lopez, I.; Lu, J.~X.;
  Iott, B.~E.; Baylin, A.; Jansen, E.~C.; Clarke, P.; et~al. 2020.
\newblock Uncovering the relationship between food-related discussion on
  Twitter and neighborhood characteristics.
\newblock \emph{Journal of the American Medical Informatics Association},
  27(2): 254--264.

\bibitem[{Yang et~al.(2021)Yang, Lin, Fang, and Zhu}]{yang2021eating}
Yang, G.-y.; Lin, X.-l.; Fang, A.-p.; and Zhu, H.-l. 2021.
\newblock Eating habits and lifestyles during the initial stage of the COVID-19
  lockdown in China: A Cross-Sectional Study.
\newblock \emph{Nutrients}, 13(3): 970.

\bibitem[{YouTube(2022)}]{YouTubepolicy}
YouTube. 2022.
\newblock YouTube API Services - Developer Policies.
\newblock https://developers.google.com/youtube/terms/developer-policies.
\newblock (accessed Jan 6, 2022).

\end{thebibliography}
}


\end{document}